\documentclass[a4paper,11pt]{article}
\usepackage{pos}

\newcommand{\NTIN}[5]{\emph{#1},\href{http://dx.doi.org/#5}{\textbf{#2}, #4 (#3)}}
\usepackage{graphicx} 
\usepackage{subcaption}
\newcommand{\etal}{\textit{et al}.}

\title{A web application for monitoring cosmic rays and solar activity}
 \ShortTitle{HVO}

\author[a]{David Pelosi}
\author[a]{Nicola Tomassetti}
\author[b]{Matteo Duranti}

\affiliation[a]{Department of Physics and Earth's Science, Universit\`a degli Studi di Perugia, Italy}
\affiliation[b]{INFN - Sezione di Perugia - Perugia, Italy}
\emailAdd{david.pelosi@studenti.unipg.it}
\emailAdd{nicola.tomassetti@unipg.it}

\abstract{The flux of cosmic rays in the heliosphere is subjected to variations that are related to the Sun's magnetic activity. To study this effect, updated time series of multichannel observations are needed. Here we present a web application that collects real-time data on solar activity proxies, interplanetary plasma parameters, and charged cosmic-ray data. The data are automatically retrieved on daily basis from several space missions or observatories. With this application, the data can be visualized and download into a common format. Along with observational data, the application aims to provide real-time calculations for the solar modulation of cosmic rays in the heliosphere.}

\FullConference{37$^{\rm{th}}$ International Cosmic Ray Conference (ICRC 2021)\\
		July 12th -- 23rd, 2021\\
		Online -- Berlin, Germany}

\begin{document}
\maketitle

\section{Introduction} 
%
During their motion inside the heliosphere, cosmic rays (CRs) interact with the heliospheric plasma. Specifically, the solar wind and its embedded magnetic field reshape their energy spectra. The resulting effect, known as solar modulation, is caused by basic transport processes such as diffusion, drift, reacceleration, convection, and adiabatic cooling \cite{ref:Potgieter2013}. Due to solar modulation, the energy spectrum of CRs observed near-Earth is considerably different from that in the surrounding interstellar medium, the so-called Local Interstellar Spectrum (LIS). 
Solar modulation is energy- and time-dependent. The modification is larger for CRs at low energies (e.g. with kinetic energies below a few GeV) and shows a quasi-peridical time variation which reflects a clear correlation with the 11-year cycle of solar activity. 
To monitor solar activity, a widely used proxy is the SunSpot Number (SSN), i.e. the number of the observed dark spots in the Sun's photosphere. The SSN is widely used as a good proxy for solar activity, varies with a period of 11 years, known as solar cycle. 
Investigating the solar modulation phenomenon is of crucial importance 
to achieve a deep understanding of the dynamics of charged particles in the heliosphere, as well as to reliably predict the radiation dose received by electronics and astronauts in space missions. 
Given the ever-growing number of satellites orbiting around Earth and human space missions to Moon and Mars planned in the next decades, forecasting the level of CR radiation in low-Earth or deep-space orbits is of great importance. For this purpose, 
several predicting models for the solar modulation of CRs have been developed\,\cite{ref:Potgieter2013,ref:FF,ref:NTPRD,ref:NTAPJ,ref:NTPRL}.
Recent CR data collected in space include the time-resolved flux measurements of the experiments EPHIN/SOHO (since 1995 to 2018) \cite{ref:EPHINSOHO}, PAMELA (2006-2016) \cite{ref:PAMELA}, and AMS-02 (since 2011 and still operative for all ISS lifetime) \cite{ref:AMS}, 
along with the direct LIS data from the Voyager probes in the interstellar space \cite{ref:VOYAGER1}.
The temporal variations of CRs fluxes are also measured indirectly by Neutron Monitors (NMs) \cite{ref:NM}.
The counting rate $N_{j}$ of a NM detector associated with $k$-type CR particles is 
defined as an integration of the total flux $J_{k}(t,R)$, above the local geomagnetic cutoff rigidity $R_{C}$,  convoluted with the yield function of the detector $Y_{k}=Y_{k}(t,R)$. The total NM rate is then obtained by the sum of the contributing CR species.
Models also need heliospheric and solar data such as, in particular, the SSN as function of time, which is provided by the SILSO/SIDC database of the \textit{Royal Observatory of Belgium}~\cite{ref:SILSO}; the strength of the polar magnetic field and the tilt angle of the heliospheric current sheet, provided on 10-day basis by the \textit{Wilcox Solar Observatory}~\cite{ref:WILCOX}. Also important are interplanetary data about the solar wind, such as its radial speed and the density of its proton component, that are distributed by NASA missions WIND and ACE~\cite{ref:NASA}.

The \textit{Heliophysics Virtual Observatory} (HVO) has been developed to make data an easy and quick access to all these data involving charged radiation, solar activity, and interplanetary plasma. 
HVO is a web application that collects all the observational data mentioned above into a unique platform which is updated automatically on daily basis. The HVO functionalities include the possibility of visualizing, manipulating and downloading updated data. We also present a simplified real-time model of near-Earth proton flux integrated into a specific section of HVO.

\section{Real-time model} 
%
The transport of CRs in the heliosphere is governed by the Parker equation: 
\begin{equation}
\label{eq:1}
\frac{\partial f}{\partial t}  =  - (C \vec V + \langle \vec v_{drift} \rangle) \cdot \nabla f + \nabla \cdot ( \textbf{K} \cdot \nabla f) + \frac{1}{3} (\nabla \cdot \vec V)  \frac{\partial f}{\partial \ln R} + q
\end{equation}  
The equation describes the temporal evolution of CR phase space density $f=f(t,R)$, where $R=p/Z$ is the magnetic rigidity of CRs, $\langle \vec v_{drift} \rangle$ is their averaged drift velocity, $\vec V$ is the solar wind velocity, $\textbf{K}$ is the symmetric part of CR diffusion tensor, and $q$ is any local source of CRs \cite{ref:Potgieter2013}. 
The Parker equation is often resolved in under simplifying approximation, for example in the so-called Force-Field (FF) model~\cite{ref:FF}. 
The FF model assumes steady-state conditions (i.e. negligible short-term modulation effects), 
radially expanding wind $V(r)$, an isotropic and separable diffusion coefficient $K\equiv\kappa_{1}(r){\cdot}\kappa_{2}(R)$, negligible losses and no drift. 
In spite of its questionable assumptions, the FF model is widely used 
as it provides a simple and practical way to describe the near-Earth CR flux and its long-term evolution.
From Eq.\,\ref{eq:1}, the CR flux $J(t,R)$ is given by $J=R^{2}f$. In terms of kinetic energy per nucleon $E$, for a CR nucleus with charge number $Z$ and mass number $A$, the FF equation for the near-Earth ($r=1$\,AU) flux at the epoch $t$ is given by: 
\begin{equation}
\label{eq:2}
J(E,t) = \frac{(E + M_p)^2  - M_p^2}{(E + M_p +\frac{Z}{A}\phi(t)) ^2 -M_p^2}  J_{\rm{LIS}} (E+\frac{Z}{A}\phi(t)),
\end{equation}
where the parameter $\phi$ is the so-called \emph{modulation potential}. It has the units of an electric potential and lies in the typical range 100-1000 MV. The parameter $\phi$ can be interpreted as the averaged energy loss per charge units of CR particles in the heliosphere.
The implementation of a CR modulation model under the FF approximation depends on the knowledge of two key elements: the time-series of the $\phi$ parameters, and the input LIS model $J=J(E)$ describing the energy spectrum of CRs outside the heliosphere.
In this work, we have used new LIS models based on the latest results from Voyager 1 and AMS-02 \cite{ref:LIS} and the values of the modulation potential reconstructed by 
\textit{Usoskin et al.\,2011}\,\cite{ref:USOSKIN}, from NM data on monthly basis, since 1964 to 2011. %
To set up the $\phi$ reconstruction after 2011 to the present epoch, instead of repeating the Usoskin methodology (based on the calibration of NMs and the evaluation of their yield function), we have adopted a simplified method. For a given NM detector, the NM counting rate $N(t)$ and the parameter $\phi(t)$ are well anti-correlated and we can establish a quadratic relation between them:
\begin{equation}
\label{eq:3}
\phi(N(t)) = A + B \cdot N(t) + C \cdot N(t)^2
\end{equation}
We determined the coefficient $A$, $B$, and $C$ as best-fit values using NM counts of many stations, analyzed against the $\phi$-parameter from \textit{Usoskin et al.\,2011}\,\cite{ref:USOSKIN}.
This approach enabled us to obtain a prediction of the parameter $\phi$ for any epoch $t$ for which the NM rate $N$ is known.
Inserting  the parametric model of LIS and the $\phi$ value at epoch $t$ in Eq.\,\ref{eq:2}, 
we obtain a real-time evaluation of the near-Earth CR flux, based on FF calculations calibrated against NM data. This simple model has been integrated into HVO. 
\begin{figure}[h!]
\centering
    \begin{subfigure}[t]{0.43\textwidth}
        \includegraphics[width=\linewidth]{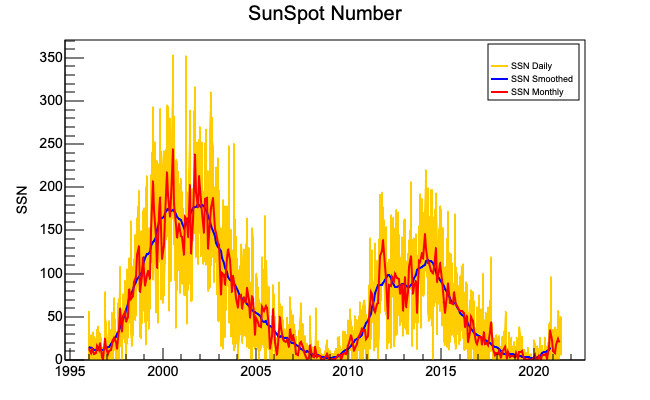}
        \caption{Daily, monthly and smoothed SSN in the time interval 1/1/1996 - 23/6/2021.}
        \label{fig:PC}   \vspace*{2mm}
    \end{subfigure}\hspace{1em}%
    \begin{subfigure}[t]{0.43\textwidth}
        \includegraphics[width=\linewidth]{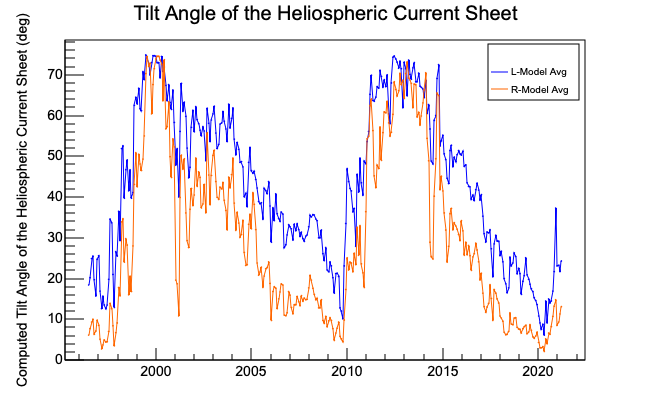}
        \caption{Averaged tilt angle of the HCS measured with the classic and radial model for any Carrington rotation between 1/1/1996 and 23/6/2021.}
        \label{fig:PA}    \vspace*{2mm}
    \end{subfigure}
    \begin{subfigure}[t]{0.43\textwidth}
        \includegraphics[width=\linewidth]{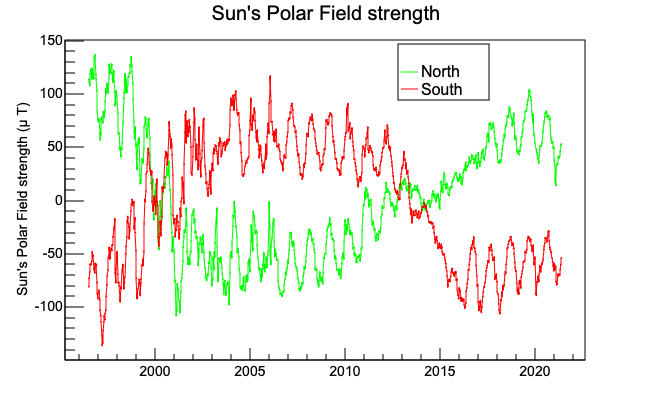}
        \caption{Sun's Polar Field strength (SFS) for northern and southern hemisphere in the time interval 1/1/1996 - 23/6/2021.}
        \label{fig:PS}    \vspace*{2mm}
    \end{subfigure}\hspace{1em}%
    \begin{subfigure}[t]{0.43\textwidth}
        \includegraphics[width=\linewidth]{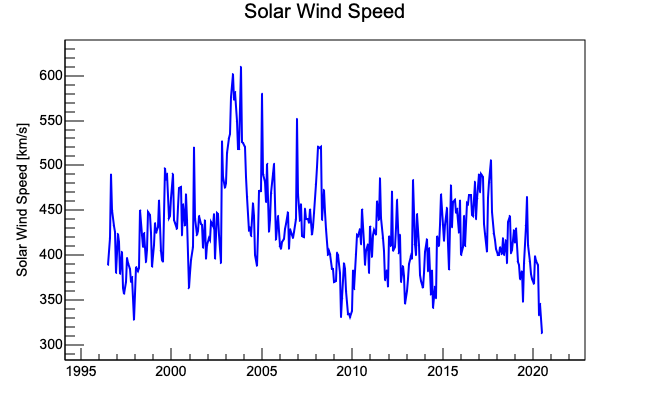}
        \caption{Radial speed of solar wind for the period 1/1/1996 - 23/6/2021.}
        \label{fig:PI}   \vspace*{2mm}
    \end{subfigure}
    
     \begin{subfigure}[t]{0.43\textwidth}
        \includegraphics[width=\linewidth]{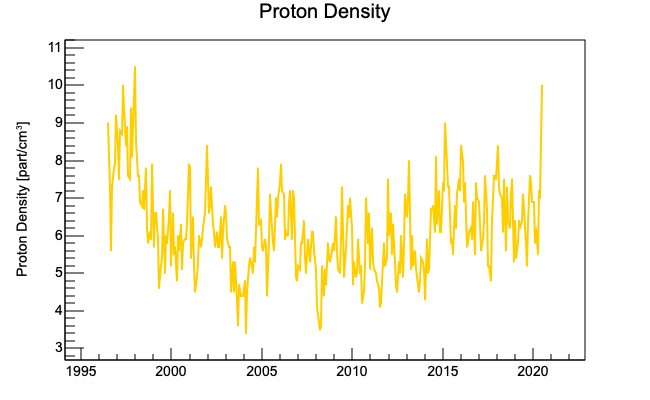}
        \caption{Proton density of solar wind for the period 1/1/1996 - 23/6/2021.}
        \label{fig:PS}   \vspace*{2mm}
    \end{subfigure}\hspace{1em}%
    \begin{subfigure}[t]{0.43\textwidth}
        \includegraphics[width=\linewidth]{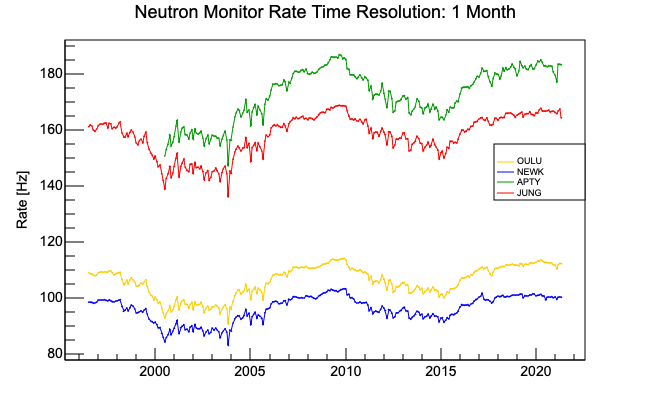}
        \caption{Monthly rate $N(t)$ of the NM stations: OULU, NEWK, APTY, JUNG. Time interval 1/1/2004 - 1/1/2021.}
        \label{fig:PI}    \vspace*{2mm}
    \end{subfigure}
    
    \caption{Data available on HVO.}
    \label{fig1}
\end{figure}

\begin{figure}[h!]
  \centering
  \begin{subfigure}[t]{0.6\textwidth}
    \includegraphics[width=\linewidth]{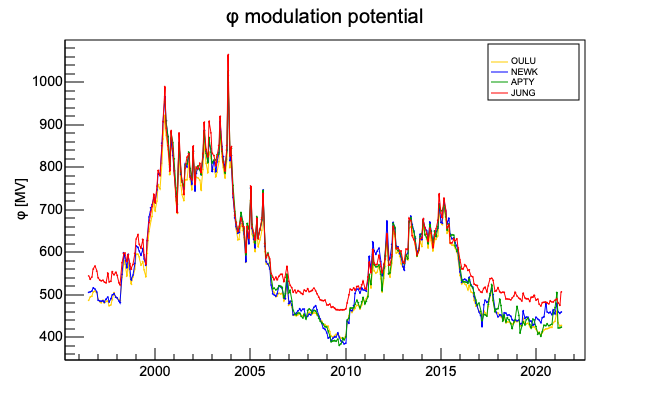}
    \caption{Calculated time-series of $\phi$ (1/1/1996 - 23/6/2021) using the rates of NM stations: OULU, NEWK, APTY, JUNG.}
  \vspace*{5mm}
  \end{subfigure}
  
  \begin{subfigure}[t]{0.6\textwidth}
    \includegraphics[width=\linewidth]{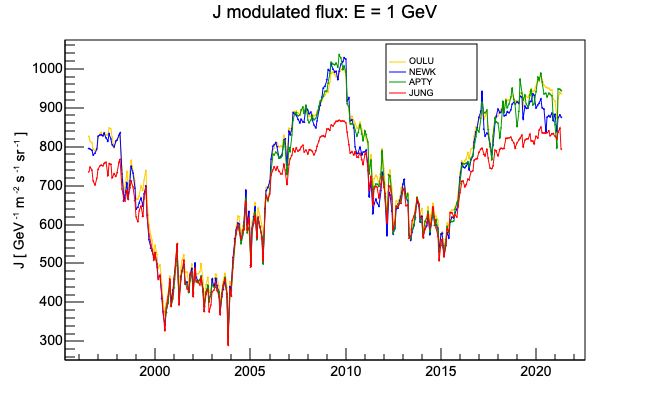}
    \caption{Estimated near-Earth proton flux at $E$ = 1 GeV using the time-series of $\phi$ reconstructed from the rates of NM stations: OULU, NEWK, APTY, JUNG.}
  \end{subfigure}
 \caption{Real-time flux model in HVO}
 \label{fig2}
\end{figure}

\section{The Heliophysics Virtual Observatory}
The investigation of the solar modulation phenomenon requires a large variety of heliospheric and radiation data. HVO \cite{ref:HVO} is a project developed 
under the CRISP scientific program of experimental study and phenomenological modeling of space weather, within the framework agreement between \textit{Università degli Studi di Perugia} and \textit{Agenzia Spaziale Italiana} (ASI).
HVO is a web application that daily extracts data with Python scripts from several databases (listed in Resources section). It visualizes and makes them available in a standardized format. 
HVO is implemented with the JavaScript \texttt{ROOT} package \texttt{JSROOT}. The application allows users to manipulate data graphs directly from the web, to download data as machine readable text, graphic objects in \texttt{ROOT} format or PNG images. HVO has is organized in three main sections. The first one is dedicated to solar data such as SSN in daily, monthly, yearly, and smoothed formats extracted from the SILSO/SIDC data center\,\cite{ref:SILSO}, observations of the Sun's polar magnetic field strength and tilt angle of the HCS reconstructed with the classic and radial model from the Wilcox Solar observatory~\cite{ref:WILCOX}. The second section contains heliospheric data such as the radial speed of the solar wind and the density of its proton component, updated on monthly bases from WIND and ACE missions\,\cite{ref:NASA}. The third section contains cosmic radiation data from several NM stations and a real-time model for the flux of CR protons, as presented in Sect.\,2. An interactive user interface provides the possibility to select one or more NM stations, to choose the time resolution of the rates (daily, monthly, yearly and by Carrington rotation), set the proton CR energy and time range.
For each selected NM station, HVO provides the graph of the counting rate $N(t)$, the corresponding calculation of the modulation potential $\phi(t)$ (from Eq.\,\ref{eq:3}), and the estimated near-Earth flux of CR protons $J(t,E)$ (from Eq.\,\ref{eq:2}).
An example of the HVO functionalities is shown in Fig.\,\ref{fig1} and Fig.\,\ref{fig2}.

\section{Conclusions}
We have presented a web application aimed at monitoring solar activity and cosmic radiation, as well as providing real-time calculation of the energy spectra of CRs in proximity of the Earth.
HVO is a useful tool for the CR astrophysics and space physics community. 
Future development may include an improvement of the real-time CR flux model, its extension the to other charged species and other locations in the interplanetary space. 
In this respect, HVO can also integrated with state of the art numerical models of CRs transport in the heliosphere\,\cite{ref:NTPRD,ref:NTAPJ}, that will enable us to forecast the CR radiation at an interplanetary level. 
Other extension of HVO may include space weather data, e.g. on the occurrence of solar flares and coronal mass ejections, geomagnetic storms, or other interplanetary disturbance phenomena.
\acknowledgments
We acknowledge the support of ASI under agreement ASI-UniPG 2019-2-HH.0.

\end{document}